\documentclass[aps,prl,twocolumn,superscriptaddress]{revtex4-1}
\usepackage{graphicx}
\usepackage{bbold}
\usepackage{amsmath}
\usepackage{upgreek}
\usepackage{xcolor}
\newcommand{\ket}[1]{{\left| {#1} \right\rangle}}

\newcommand{\ketbra}[2]{{\left| {#1} \right\rangle
    \!\!\left\langle{#2} \right|}}

\DeclareMathAlphabet{\mathbsf}{OT1}{cmss}{bx}{n}

\begin{document}
\title{Laserless quantum gates for electric dipoles in thermal motion}
\author{Eric R. Hudson and Wesley C. Campbell}
\affiliation{Department of Physics and Astronomy, Los Angeles, California 90095, USA}
\affiliation{UCLA Center for Quantum Science and Engineering, University of California – Los Angeles, Los Angeles, California 90095, USA}

\begin{abstract}
Internal states of polar molecules can be controlled by microwave-frequency electric dipole transitions. 
If the applied microwave electric field has a spatial gradient, these transitions also affect the motion of these dipolar particles.
This capability can be used to engineer phonon-mediated quantum gates between e.g. trapped polar molecular ion qubits without laser illumination and without the need for cooling near the motional ground state.
The result is a high-speed quantum processing toolbox for dipoles in thermal motion that combines the precision microwave control of solid-state qubits with the long coherence times of trapped ion qubits. 
\end{abstract}

\date{\today}
\maketitle

Trapped atomic ion qubits have demonstrated the highest-fidelity quantum operations of all systems~\cite{Christensen2019,Ballance16,Harty14,Gaebler16}, yet challenges remain for their integration into large, scalable platforms. 
These systems typically rely on laser-driven, phonon-mediated quantum gates, which introduce three issues for producing large-scale devices.
First, the production, conditioning, and delivery of the requisite laser light is not yet available as integrable subsystems.
Second, laser-induced spontaneous scattering from qubits during gate operations limits the achievable gate fidelity~\cite{Ozeri2007}, which sets the number of physical qubits to achieve a fault-tolerant logical qubit.
Third, these phonon-mediated gates typically require cooling the ions to near the ground state of motion, \textit{i.e.}~to the Lamb-Dicke regime, where the spatial extent of the motional state is much smaller than the wavelength of the laser.
This adds technical complexity and renders the gate fidelity susceptible to corruption by the heating of the motional modes from nearby surfaces~\cite{Sedlacek2018}.

At present, several potential solutions to these challenges are being pursued.
Integrated photonics are being developed that could provide a scalable means to deliver the requisite lasers~\cite{Bruzewicz2019,Karan2020}, if they can be extended to handle the intensity and short wavelengths necessary for atomic ion qubits~\cite{Mehta16}. 
Schemes for laserless gates for atomic ion qubits are likewise under development that use magnetic field gradients~\cite{Mintert2001,Johanning2009,Khromova2012,Ospelkaus2008,Ospelkaus2011,Brown2011,Timoney2011} to couple the internal degrees of freedom to ion motion.  
These protocols are typically slower than their laser-based counterparts by a factor of
the fine structure constant $\alpha$, as they rely on an atomic-scale \emph{magnetic} (as opposed to \emph{electric}) dipole interaction. 
Last, `ultrafast' gate schemes have been developed, based on state-dependent forces generated by lasers~\cite{GarciaRipoll2003,Duan2004} or magnetic field gradients~\cite{MurPetit2012TemperatureIndependent}, that can in principle operate outside of the Lamb-Dicke limit.

Here, we consider an alternative route to scalable, trapped ion quantum information processing that uses \emph{electric-field gradients} produced by multipole electrode configurations, including the trap itself, to couple the internal states of polar molecular ions to collective phonon modes of their Coulomb crystal -- in principle this interaction could also be used with Rydberg ions~\cite{Zhang2020}.
As the electrodes dictate the shape of the applied microwave electric field, this coupling can be made to be independent of the ion position and therefore largely independent of the motional state. 
As such, these electric-field gradient gates (EGGs) comprise a toolbox that provides fast state preparation and measurement (SPAM), as well as  single- and two-qubit gate capabilities for ions \textit{in thermal motion}.
EGGs therefore have inherent advantages for scaling, as they combine the precision microwave qubit control enjoyed by solid-state qubits with the long coherence time of trapped ion qubits. 
In what follows, we introduce the basic EGGs interaction and show how it can be used for the necessary quantum logic operations. 

\begin{figure}
    \centering
    \includegraphics[width = 0.49\textwidth]{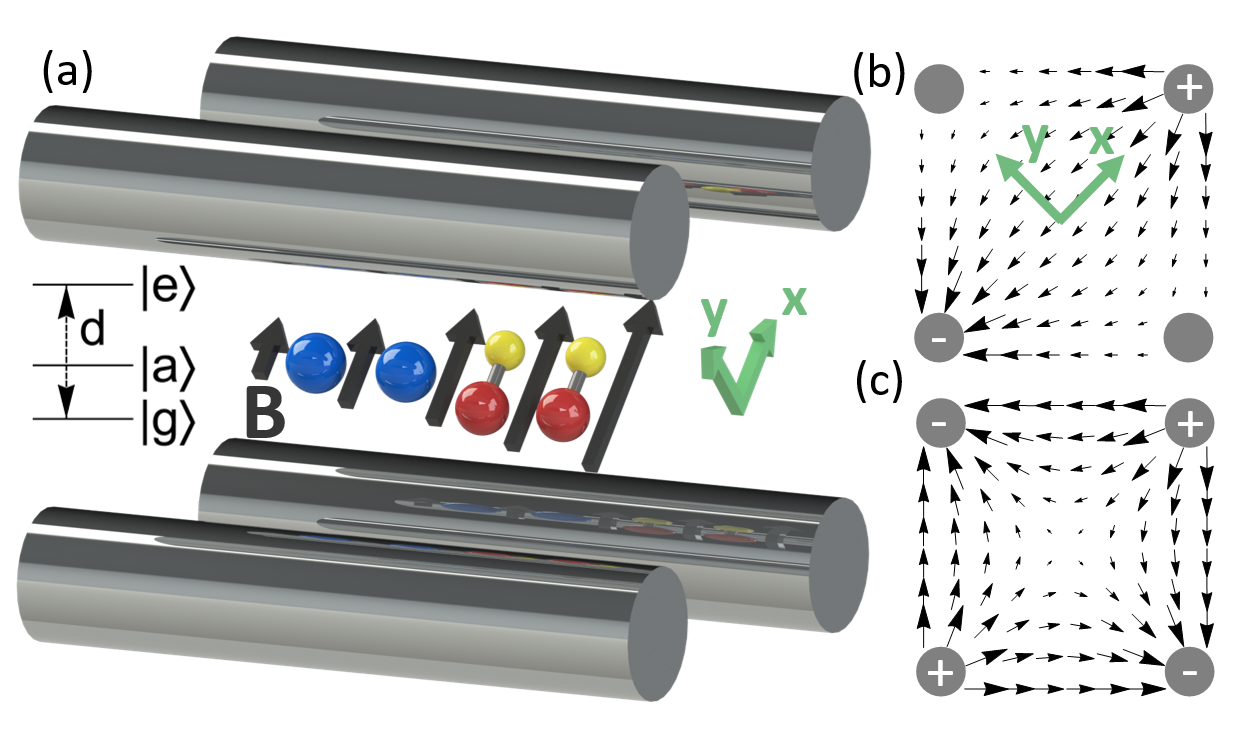}
    \caption{(a) Schematic of the basic EGGs system. A spatially varying magnetic field provides individual molecule addressability. Doppler cooled co-trapped atomic ions provide sympathetic cooling and molecular state-readout capabilities. Applying microwave voltages in a (b) dipole configuration allows single qubit gates, while a (c) quadrupole configuration provides SPAM and two-qubit gates as described in the text.
    }
    \label{fig:setup}
\end{figure}

To illustrate EGGs, we consider a linear ion chain that may contain both atomic and molecular ions of approximately the same mass (for simplicity, we assume they are equal), and primarily consider motion along one radial ($x$) direction of the chain, see Fig.~\ref{fig:setup}.  
In a linear Paul trap, harmonic confinement in the radial directions is provided by a time-dependent electric potential of the form $\Phi(\mathbf{r},t) = V_o\cos(\Omega_\mathrm{rf} t)\left(x^2 - y^2\right)/r_o^2$, where $V_o$ is the amplitude of the radio-frequency voltage applied to the trap electrodes at frequency $\Omega_\mathrm{rf}$ and $r_o$ is the trap field radius. 
This provides a ponderomotive potential leading to ion motion that can be approximated by the Hamiltonian $\mathcal{H}_o/\hbar = \sum_p\omega_p (a_p^\dagger a_p + \frac{1}{2})$, where $\omega_p$ is the frequency of normal mode $p$. 
The displacement of ion $i$ from its equilibrium position can be written as a superposition of displacements of the normal modes:
$\hat{x}^{(i)} = \sum_p  \sqrt{\hbar/(2 m \omega_p)} \mathrm{b}_p^{(i)}(a_p + a_p^\dagger)$~\cite{James1998}.

Molecular ions in the chain are assumed to be identical polar molecules, each with a pair of opposite-parity states $\ket{g^{(i)}}$ and $\ket{e^{(i)}}$ that represent the -1 and +1 eigenstates, respectively, of the Pauli operator $\sigma_Z^{(i)}$ for this effective two-level system of molecule $i$.  Further, each molecule will also possess long-lived, magnetic field insensitive auxiliary states $\ket{a^{(i)}}$ that can be used for shelving and information storage. 
The molecules are subject to a static magnetic field $B(z)\mathbf{\hat{x}}$ that defines the quantization axis and whose magnitude has a gradient along $\mathbf{z}$. $\ket{g^{(i)}}$ and $\ket{e^{(i)}}$ represent states with the same total angular momentum projection, $m_F$, along $\mathbf{x}$.  
The qubit states are separated in energy by the noninteracting Hamiltonian $\mathcal{H}_\mathrm{mol}^{(i)}/\hbar = (\Delta^{(i)}/2)\sigma_Z^{(i)}$, with the qubit states chosen such that $\Delta^{(i)}$ is in the radio- or microwave-frequency range (i.e.~$\Delta^{(i)} \gg \omega_p$, but still low enough that precision control technology is readily available).
These states are connected by an electric dipole transition moment according to $d = \langle e|\mathbf{d}|g\rangle \cdot \mathbf{\hat{x}}$.
Physically, these states could be any dipole-connected states, such as rotational states or $\Omega$-, $l$-, or $K$-doublets, and they define the Hilbert space of the dipole.

Transitions between the molecular states can be driven by applying a sinusoidal voltage of frequency $\omega_\mathrm{m}$ and amplitude $V_\mathrm{m}$ to the Paul trap electrodes to produce an electric field that interacts with the dipole according to $\mathcal{H}_\mathrm{m} = -\mathbf{d}\cdot \mathbf{E}_\mathrm{m} (\mathbf{r})$. 
If the electrodes are driven in a dipole configuration (Fig.~\ref{fig:setup}(b)), the electric field due to time-dependent voltage at the position of the ions is $\mathbf{E}_{E1} \cdot \mathbf{\hat{x}} \approx -V_\mathrm{m}\cos(\omega_\mathrm{m} t + \phi_\mathrm{m})/(2 r_o)$.
Therefore, transitions driven in this manner are described in the interaction picture with respect to $\mathcal{H}_\mathrm{mol}^{(i)}$ by
\begin{align}
    \mathcal{H}_{E1}^{(i)}/\hbar = -d\,\, \mathbf{\hat{x}} \cdot \mathbf{E}_{E1}/\hbar = \frac{\Omega}{2}\left(\sigma_{+}^{(i)} e^{\imath\delta^{(i)} t} + \textrm{h.c.}\right)
\end{align}
where  $\Omega = d V_\mathrm{m}/(2 r_o \hbar)$, $\delta^{(i)} = \Delta^{(i)} - \omega_\mathrm{m}$, $\sigma_{+}^{(i)}$ is the Pauli raising operator for the $\{\ket{e^{(i)}},\ket{g^{(i)}}\}$ subspace, and $\mathrm{h.c}$ denotes the Hermitian conjugate.  We have assumed the microwave phase can be taken to be $\phi_\mathrm{m} = 0$, and the rotating wave approximation (RWA) has been used to eliminate the counter-rotating terms. 

If, on the other hand, the trap electrodes are driven in a quadrupole configuration (Fig.~\ref{fig:setup}(c)), the electric field due to time-dependent voltage at the position of the ions is
$\mathbf{E}_{E2}\cdot\hat{\mathbf{x}} = -2V_\mathrm{m}x\cos(\omega_\mathrm{m} t + \phi_\mathrm{m})/r_o^2$. 
Therefore, transitions driven in this manner are described in the interaction picture with respect to $\mathcal{H}_\mathrm{mol}^{(i)}$ by 
\begin{align}
    \mathcal{H}_{E2}^{(i)}/\hbar =&  -d\,\, \mathbf{\hat{x}} \cdot \mathbf{E}_{E2}/\hbar \nonumber \\
    =&\frac{2\Omega}{r_o} \sum_p  \sqrt{\frac{\hbar}{2 m \omega_p}} \mathrm{b}_p^{(i)}(a_p + a_p^\dagger)\left(\sigma_{+}^{(i)}e^{\imath\delta^{(i)} t}+\textrm{h.c.}\right) \nonumber\\
    & + x_\mathrm{eq} \frac{2\Omega}{r_o}\left(\sigma_{+}^{(i)} e^{\imath\delta^{(i)} t} + \textrm{h.c.}\right),
    \label{eq:HE2}
\end{align}
where the RWA has been used to eliminate terms that oscillate at $\Delta^{(i)} + \omega_\mathrm{m}$. 
Here, $x_\mathrm{eq}$ is the equilibrium $x$ position of the trapped ion,  which may differ from the microwave field null (due to, \textit{e.g.}~stray static electric fields).
For $x_\mathrm{eq} = 0$, $\mathcal{H}_{E2}^{(i)}$ can only drive `sideband' transitions that couple differing molecular states while creating or destroying a single phonon in a mode $p$, while `carrier' transitions which couple differing molecular states without changing the state of the phonons can be driven by $\mathcal{H}_{E1}^{(i)}$. 
If $x_\mathrm{eq} \neq 0$, $\mathcal{H}_{E2}^{(i)}$ may also drive carrier transitions.
In what follows, we assume $x_\mathrm{eq} = 0$ unless otherwise noted.

Further, because the electric field is divergenceless, the gradient along $\mathbf{\hat{x}}$ is accompanied by gradients along $\mathbf{\hat{y}}$ and/or $\mathbf{\hat{z}}$.
In principle, these gradients can drive transitions that change $m_F$ by $\pm 1$. However, these transitions experience a Zeeman shift of order $\mu_B B$ ($\mu_B$ is the Bohr magneton), and we assume $B$ is large enough that their effect can be neglected. 
Together, $\mathcal{H}_{E1}$ and $\mathcal{H}_{E2}$ provide a complete set of tools for quantum logic with trapped polar molecular ions that, as we show below, does not require ground-state cooling.

For concreteness, we consider EGGs operations in $^{29}$Si$^{16}$O$^+$ in a trap with a $\omega_1 = 2\pi\cdot 1$~MHz radial center of mass secular frequency and $r_o = 0.5$~mm.
This molecular ion, which has a dipole moment of $d \approx (4 \mbox{ D})/\sqrt{3}= 2.3$~D \cite{Cai1998AbInitio}, is particularly attractive as $^{28}$Si$^{16}$O$^+$ has recently been optically pumped into its ground rovibrational state~\cite{Stollenwerk2020Cooling} and the nuclear spin $I = 1/2$ of $^{29}$Si$^{16}$O$^+$ provides a convenient magnetic field insensitive subspace for storing quantum information, as outlined in Ref.~\cite{Hudson2018}.

\textit{State preparation and measurement.} -- Preparation and measurement of a molecular ion quantum state can be achieved using EGGs to produce a state-dependent motional excitation that is subsequently read out by a co-trapped atomic ion.  
The dipole-motion coupling is produced by simultaneously applying, in the quadrupole configuration, two microwave tones that are resonant with the first-order motional sidebands of the $\ket{e^{(i)}} \leftrightarrow \ket{g^{(i)}}$ transition for a motional mode $q$: $\omega_{\mathrm{m}}^\pm = \Delta^{(i)} \pm \omega_q$. 

For this situation, in the interaction picture with respect to $\mathcal{H}_o$ and $\mathcal{H}_\mathrm{mol}^{(i)}$ 
and neglecting the time-dependent terms,
the total Hamiltonian takes the form: 
\begin{align}
    \mathcal{H}^{(i)}_{h}/\hbar &=  2\Omega\eta_q^{(i)}(a_q + a_q^\dagger)\sigma_X^{(i)},
    \label{eq:SDH}
\end{align}
where we define $\eta_q^{(i)} \equiv \sqrt{\hbar/(2 m \omega_q r_o^2)} \mathrm{b}_q^{(i)}$.
This interaction leads to the time evolution operator
\begin{align}
    U_h^{(i)} = \ketbra{-X^{(i)}}{-X^{(i)}}\,D_q \!\left(2\imath\Omega\eta_q^{(i)} t \right) \nonumber\\
    + \ketbra{+X^{(i)}}{+X^{(i)}}\,D_q \!\left(-2\imath\Omega\eta_q^{(i)} t \right), 
    \label{eq:USDH}
\end{align}
where $\ket{\pm X^{(i)}} = \left(\ket{g^{(i)}} \pm \ket{e^{(i)}}\right)/\sqrt{2}$ and $D_q(\alpha) \equiv \exp(\alpha a_q^\dagger - \alpha^\ast a_q)$ is the harmonic oscillator displacement operator for mode $q$. 
As the average phonon number of a coherent state $\ket{\alpha}$ is $ |\alpha|^2$, this interaction adds energy, regardless of the qubit state, to the motional mode in the amount of $\Delta E_q \approx \hbar\omega_q(2\Omega\eta_q^{(i)} t)^2$.

State preparation of molecular ion qubits can then be divided into two regimes -- preparation \emph{into} the $\{\ket{g},\ket{e}\}$ qubit subspace, and preparation of pure states \emph{within} the qubit subspace.  
For the former, the interaction described by Eq.~(\ref{eq:USDH}) adds significant energy to the motional mode if the molecule is in the qubit subspace, which is heralded by monitoring a co-trapped atomic ion.
A null measurement can be followed by molecular population redistribution via blackbody radiation, spontaneous emission, or applied fields until the atomic ion confirms that the molecule is within the qubit subspace.

If the molecule is in the qubit subspace, state detection is possible by a projective quantum measurement that leaves the basis states intact, i.e. a quantum nondemolition (QND) measurement, as follows. 
Molecular ions with at least one co-trapped atomic ion can be Doppler cooled into a linear ion chain (ground state cooling is not necessary).
A molecule-specific, microwave carrier transition (\textit{i.e.}~in the dipole configuration) can transfer molecule $i$ from e.g $\ket{g^{(i)}}$ to $\ket{a^{(i)}}$. 
A Hadamard gate on the qubit subspace, accomplished via frequency-resolved microwaves driven in a dipole configuration, transfers any population in $\ket{e^{(i)}}$ to $\ket{+X^{(i)}}$.
Next, the bichromatic microwaves can be applied in the quadrupole configuration, adding energy to motional mode $q$ (Eq.~(\ref{eq:USDH})) if the ion is in $\ket{+X^{(i)}}$. 
By querying the co-trapped atomic ion (via e.g. the Doppler recooling method), the ion will be found in either $\ket{+X^{(i)}}$ or $\ket{a^{(i)}}$ (corresponding to $\ket{e^{(i)}}$ and $\ket{g^{(i)}}$, respectively).
Subsequent single qubit operations can return the ion to any desired pure state in the qubit subspace and, thus, this process can also be used for state preparation within the qubit subspace.

Since $\langle \pm X^{(i)}|\mathbf{d}|\pm X^{(i)}\rangle\cdot\mathbf{\hat{x}} = \pm d$, the state-dependent displacement effected by this bichromatic interaction can be understood as the driving of a time-varying dipole due to the force, $\mathbf{F} =$\boldmath${\nabla}$\unboldmath$(\mathbf{d}\cdot\mathbf{E})$, from the time-varying electric field gradient.  
In contrast to laser-driven motion of atomic ions, where the validity of Eq.~(\ref{eq:USDH}) quickly breaks down once the displacement becomes comparable to the optical wavelength \cite{McDonnell2007LongLived,Poschinger2010Observing}, the electric gradient force on polar molecular ions from this quadrupole electrode arrangement remains independent of ion position until the displacement becomes comparable to the trap dimensions, allowing large displacements that are easily detectable.

\begin{figure}
    \centering
    \includegraphics[width = 0.48\textwidth]{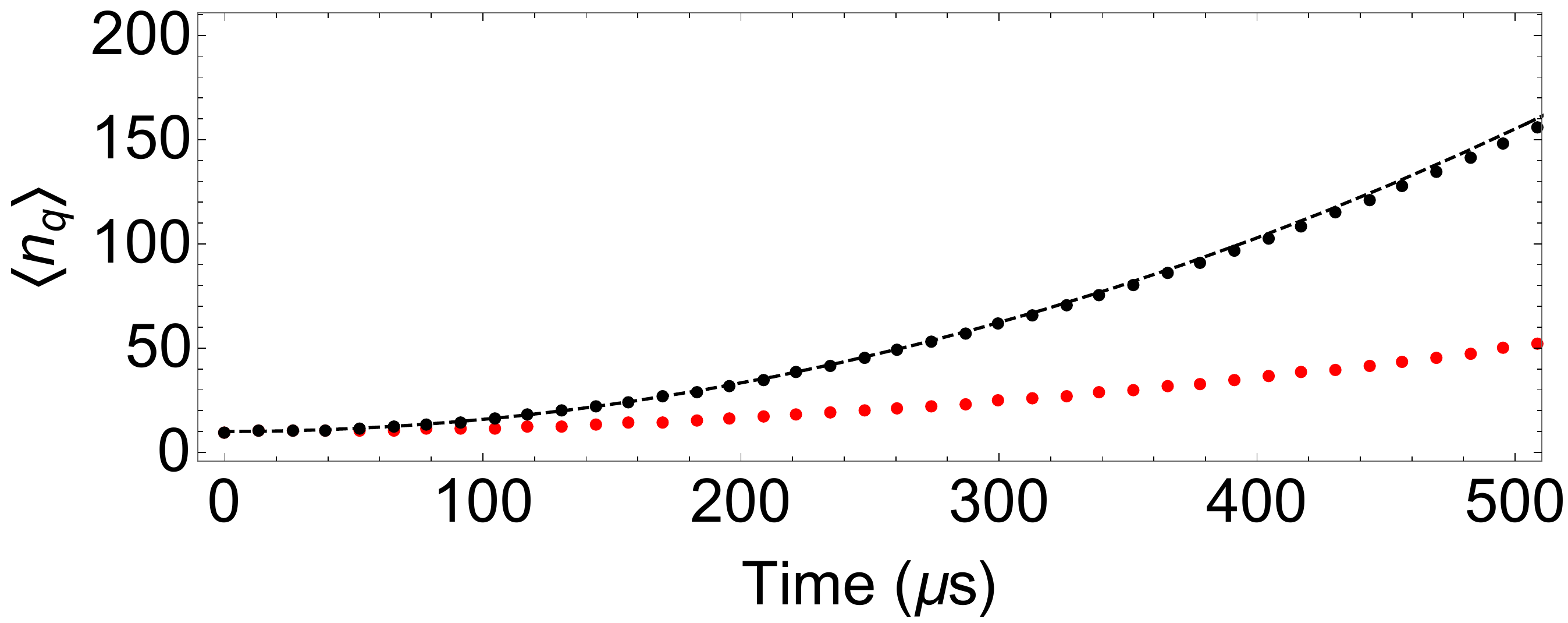}
    \includegraphics[width = 0.48\textwidth]{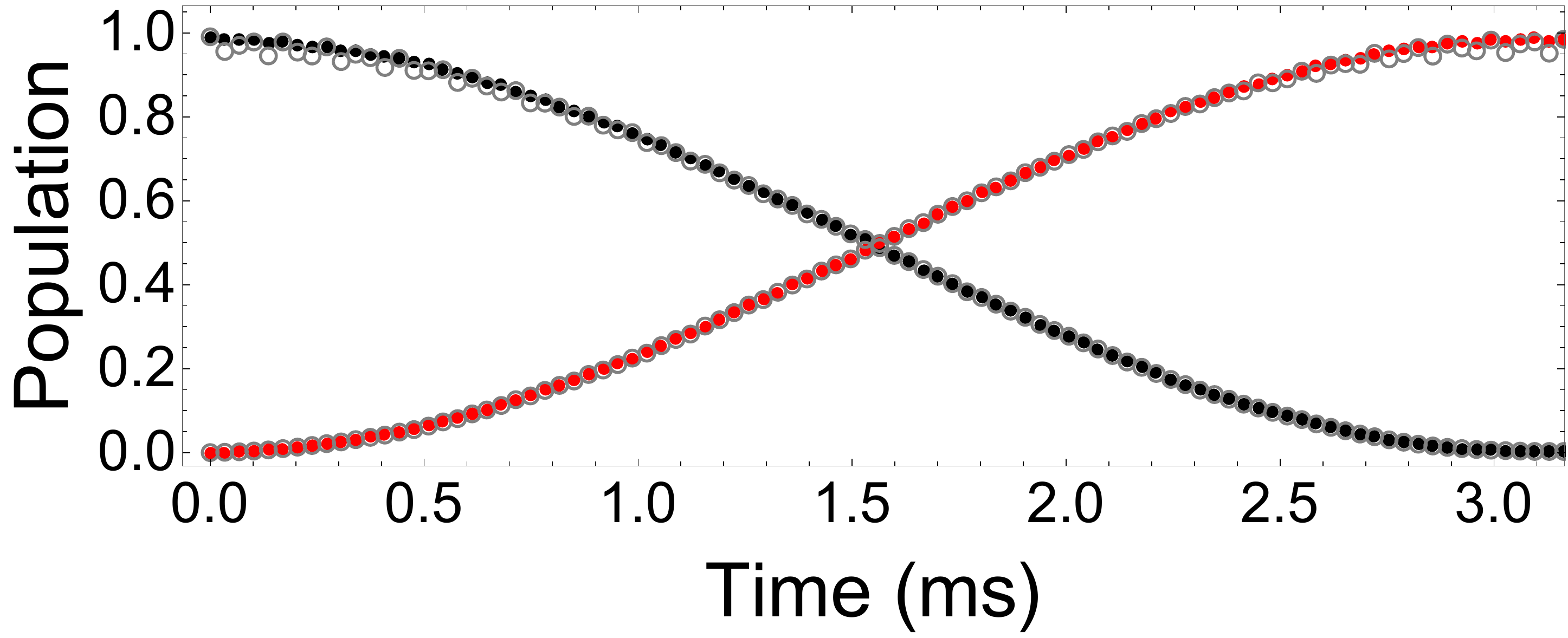}
    \caption{ 
    (a) Mean phonon number under application of microwave radiation at $\omega_\mathrm{m}^\pm = \Delta^{(i)} \pm \omega_q$ with $V_\mathrm{m} = 10$~V, $r_o = 0.5$~mm, and $\omega_q = 2\pi\cdot1$~MHz for an initial thermal state phonon distribution with $T = 0.5$~mK. 
    Black (red) dots are numerical solutions of Eq.~\ref{eq:HE2} for mode $q$ with $x_\mathrm{eq} = 0$ ($x_\mathrm{eq} = 10~\mu$m and a 5\% mismatch in sideband drive voltages), while the dashed line is the analytical result following from Eq.~\ref{eq:USDH}.
    (b) Two-qubit entangling gate based on Eq.~\ref{eq:2QGate} for a $T = 0.1$~mK thermal state.
    The black (red) points are the population in $\ket{g,g}$ ($\ket{e,e}$) as numerically determined from Eq.~\ref{eq:HE2} with $\delta = 2\pi\cdot 200$~kHz and $x_{eq} = 0$.
    The grey points are the same calculation with $x_\mathrm{eq} = 100$~nm.
    }
    \label{fig:StateDependentHeating}
\end{figure}

For easily accessible experimental parameters ($V_\mathrm{m} = 10$~V and $r_o = 0.5$~mm) the bichromatic interaction (\ref{eq:SDH}) adds roughly 150 phonons in 500~$\mu$s, shown in Fig.~\ref{fig:StateDependentHeating}(a) by the black dots and dashed line. 
If $x_\mathrm{eq} \neq 0$, carrier transitions become possible and primarily lead to an AC Stark shift.
Due to symmetry, if the motional sidebands are driven with the same amplitude, this Stark shift vanishes and the evolution is identical to the previous analytical result.
However, if there is a mismatch of the two amplitudes, a residual AC Stark shift is possible and if this shift is not accounted for, it leads to a decrease in heating rate (as shown by the red dots in Fig.~\ref{fig:StateDependentHeating}(a) for $x_\mathrm{eq} = 10~\mu$m and and mismatch of 5\%).

\textit{Single-qubit gates.} -- Following state preparation, single qubit gates on molecule $i$ are implemented by applying microwave radiation in the dipole configuration at $\omega_\mathrm{m} = \Delta^{(i)}$ and described by $\mathcal{H}_{E1}$.
Composite pulse sequences and/or shelving to $\ket{a}$ can be used to prevent unwanted phase accumulation on nearby molecules.

Interestingly, the electric-field gradient of the trap can also be used to drive carrier quadrupole transitions between, for example, rotational states separated by two rotational quanta.
The rate of this transition is roughly $(e a_o^2)\frac{2 V}{\hbar r_o^2 }\sim~1$~kHz for $V = 10$~V and $r_o = 100~\mu$m, and could be useful for e.g. shelving.
Further, additional electrodes providing higher-order multipoles could be used to drive motion on the quadrupole transition. 

\textit{Molecular two-qubit gates.} -- As the form of $\mathcal{H}_{E2}$ is similar to the Hamiltonian of an atomic ion qubit subject to a laser in the Lamb-Dicke limit, the two-qubit gates derived from that interaction can be applied to molecular ions via EGGs.
As an example, we consider a laserless version of a M{\o}lmer-S{\o}rensen gate~\cite{MolmerSorensen99Multiparticle}.
In a chain of trapped molecular ions, two ions can be isolated by shelving all other ions to $\ket{a}$ and the magnetic field gradient adjusted so that the remaining two ions have equal Zeeman shift, i.e. $\Delta^{(1)} = \Delta^{(2)} = \Delta$  
(gate operation does not require the ions have equal splittings~\cite{Inlek2017} and is assumed only for conceptual simplicity). 
Next, two microwave tones of equal amplitude at $\omega_\mathrm{m}^\pm = \Delta \pm (\omega_q' + \gamma)$ 
are applied in a quadrupole arrangement -- here $\omega_q'$ includes the microwave-field-induced AC Stark shift and $\gamma$ is the chosen detuning. The time-evolution operator for those two ions in the interaction picture with respect to $\mathcal{H}_o + \mathcal{H}_\mathrm{mol}$ and the microwave-induced Stark shift, after neglecting all time-dependent terms in the Hamiltonian, is
\begin{align}
    U = \exp\left[{-\imath \frac{2\Omega^2\eta_q^{(1)}\eta_q^{(2)}}{\gamma}\sigma_X^{(1)}\sigma_X^{(2)}t}\right].
    \label{eq:2QGate}
\end{align}
For two molecules initially in the ground qubit state, i.e.\ $\ket{g,g}$, this interaction produces a Bell state in time $t = \pi\gamma/(2\Omega^2\eta_q^{(1)}\eta_q^{(2)})$ independent of the phonon states. Taken with the SPAM and single qubit gates described above, we have shown that EGGs provides a universal gate set for trapped molecular ions. 

While the form of Eq.~(\ref{eq:2QGate}) is reminiscent of the classic M{\o}lmer-S{\o}rensen (MS) interaction often used to entangle trapped atomic ions through their interaction with a laser, the parameter regime for which Eq.~(\ref{eq:2QGate}) is valid is not the same for the two cases.
Namely, the laser-based MS interaction only takes this form in the Lamb-Dicke limit where the motional state's extent is small enough that electric field amplitude of the laser, with wavevector $\mathbf{k}$, can be accurately approximated for the purposes of the interaction as $E_o e^{-\imath \mathbf{k}\cdot\mathbf{x}} \approx E_o(1 - \imath \mathbf{k}\cdot\mathbf{x})$.
It is only in this limit, which essentially requires ground-state cooling, that the MS interaction is independent of phonon number, $n_q$.
However, for EGGs, Eq.~(\ref{eq:2QGate}) remains valid well above the ground state because the spacing of the quadrupolar electrodes sets the relevant length scale for breakdown of the approximations leading to this result.
As a result, the EGGs two-qubit gate is independent of phonon number and can be used on ions in thermal motion. 

The independence of the entangling interaction on the motional state is evident in Fig.~\ref{fig:StateDependentHeating}(b), where the evolution dictated by Eq.~(\ref{eq:2QGate}) of the states $\ket{g,g}$ (black line) and $\ket{e,e}$ (red line) are shown for a thermal state in mode $q$ with $T = 0.1$~mK, and $\gamma = 2\pi\cdot 200$~kHz.
Also, shown is the evolution of the same states for $x_\mathrm{eq} = 100$~nm (grey), where the fast oscillations in the population due to the carrier transitions are apparent (and aliased in the graph).
In principle, these are coherent and do not hamper gate fidelity, but in practice, it is likely desirable to compensate stray fields to reduce $x_\mathrm{eq}$.

$\ket{\pm X}$ are eigenstates of the symmetrically-detuned ($\gamma \! \neq \!0$) MS-type bichromatic interaction even in the presence of a carrier ($x_\mathrm{eq} \neq 0$).  As such, rotations in the $ \ket{\pm X \mp \!\! X}$ subspace are insensitive to a nonzero offset $x_\mathrm{eq}$, and the preparation of an $X$-basis Bell state from $\ket{\pm X \mp \!\! X}$ will be robust to stray uniform static fields.
For the more general case of a two-qubit quantum gate, the effect of nonzero $x_\mathrm{eq}$ on gate fidelity can be quantified by noting that, for $\ket{\pm X \pm\!\! X}$, the carrier interaction
\begin{align}
    H_\mathrm{C}/\hbar = \frac{4\Omega x_\mathrm{eq}}{r_o}(\sigma_X^{(1)}+\sigma_X^{(2)})\cos((\omega_q+\gamma) t)\label{eq:HC}
\end{align}
leads to a phase $|\Phi_\mathrm{C}| \leq 8 \Omega x_\mathrm{eq}/(r_o(\omega_q + \gamma)).$
Therefore to achieve a gate infidelity $\leq 10^{-4}$ requires $x_\mathrm{eq} \leq \hbar(\omega_q + \gamma)r_o^2/(200\sqrt{2} V_\mathrm{m} d)$, which is satisfied for modest parameters.

\textit{Ultrafast gates.} -- The EGGs interaction also provides the ability to perform entangling operations similar to the co-called ``ultrafast'' quantum gates that were developed for atomic ion systems~\cite{GarciaRipoll2003, Duan2004}. 
However, because the EGGs interaction derives its mechanical effect from a dipole interacting with a classical, continuous electric field gradient, as opposed to the discrete photon recoil so far used in the atomic ion case, its magnitude and direction are simpler to control.
If the microwave gradient is applied on resonance for a time that is shorter than $2\pi/\omega_p$ for all $p$ but long compared to $\mu_B B$ (and $2 \pi / \Delta$), then the evolution of the trapped molecular ions during that time is given solely by $\mathcal{H}_{E2}$ (Eq.~(\ref{eq:HE2})).
For two molecular ions, such a microwave pulse leads to time evolution given by:
\begin{align}
   U_p =& \,\,\,\,\,\,\,|\!- \! X-\! \! X\rangle\langle -X-\! \! X| \,D_1(2\imath\Delta p_1)\nonumber\\
    &+ |\!- \! X+ \!\! X\rangle\langle -X +\!\! X| \, D_2(2\imath\Delta p_2)\nonumber\\
    &+ |\!+ \!X -\!\! X\rangle\langle +X-\! \!X|\, D_2(-2\imath\Delta p_2)\nonumber\\
    &+ |\!+ \!X+ \! \!X\rangle\langle +X+\!\! X|\, D_1(-2\imath\Delta p_1),
\end{align}
where $\Delta p_p = \Omega\eta_p t$ with $\eta_p = |\eta_p^{(i)}|$. 
\begin{figure}
    \centering
    \includegraphics[width = 0.48\textwidth]{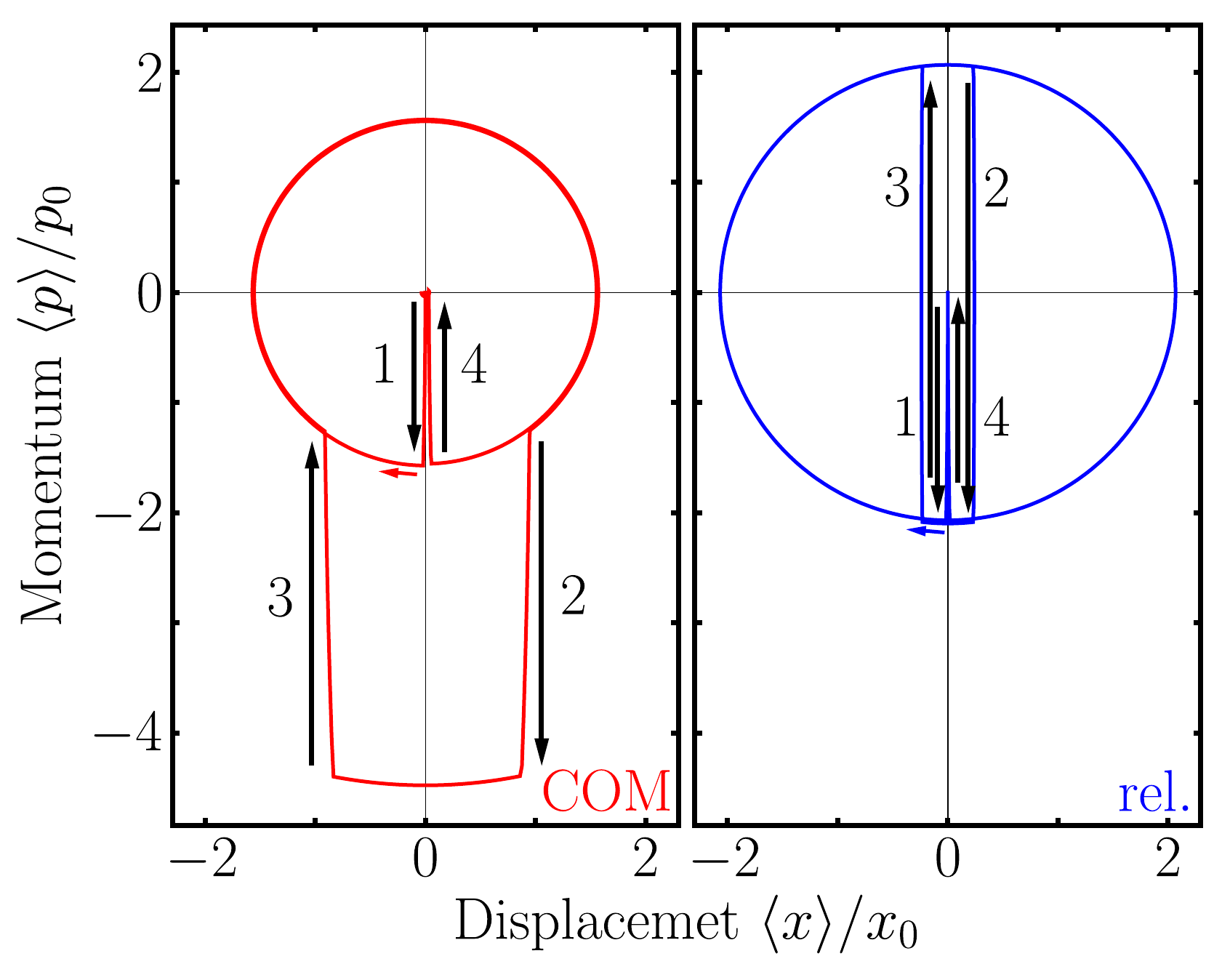}
    \caption{The phase space trajectory for two ions in the states $\ket{+X+\!X}$ (red) and $\ket{+X-\!X}$ (blue).  $x_0 \equiv \sqrt{\hbar/2m\omega_p}$ and $p_0 \equiv \sqrt{\hbar m \omega_p/2}$ are the position and momentum space widths of the ground state wavefunctions in the center of mass (red) and relative (blue) modes. Arrows indicate the action of the four pulses.
    }
    \label{fig:UFG_PhaseSpace}
\end{figure}

The effect of $N$ such microwave pulses, interspersed with free evolution for time $t_\ell$ and described by  $U_o = \prod_{p} \exp\left(-\imath \omega_p a_p^\dagger a_p t_\ell\right)$, on an arbitrary coherent state $\ket{\alpha}_p$ can be found by repeated evolution according to $U_o D_p(\pm 2\imath\Delta p_{p,j})$~\cite{GarciaRipoll2003}, where $\Delta p_{p,j}$ is the momentum displacement of pulse $j$ applied at time $T_j = \sum_{\ell = 1}^{j-1} t_\ell$.
If the pulse sequence is constructed such that $\sum_{j=1}^N \Delta p_{p,j} e^{\imath \omega_p T_{j}} = 0$ for $p = 1,2$, the effect is to return the molecules to their original motional state with an accumulated state-dependent phase.
In this case, the time-evolution operator becomes

\begin{align}
    U_N =& e^{\imath\Phi \sigma_X^{(1)} \sigma_X^{(2)}} \prod_{p} e^{-\imath \omega_p a_p^\dagger a_p T_j}\label{eq:Uufeggs}\\
    \Phi =&  2\sum_{j=2}^N\sum_{k=1}^{j-1}\Delta p_{1,j} \Delta p_{1,k} 
    \Bigg(\sin\left(\omega_1 (T_j - T_k)\right) \nonumber\\
    & -  \sqrt[4]{3}\sin\left(\frac{\omega_1}{\sqrt{3}} (T_j - T_k)\right)\Bigg)\label{eq:Phiufeggs}
\end{align}
and if $\Phi = \pi/4$, this accomplishes a controlled phase gate.

This result is the same in Ref.~\cite{GarciaRipoll2003,Duan2004} (though in the $X$ basis instead of $Z$) and the pulse sequences presented by those authors are applicable for ultrafast EGGs.
Figure~\ref{fig:UFG_PhaseSpace} shows trajectories under the pulse sequence defined as `protocol 1' in Ref.~\cite{GarciaRipoll2003}. 
Here, for population in the $\ket{\pm X \pm \!X}$ subspace, the center-of-mass mode is excited (black curve), while if the molecules are in the $\ket{\pm X \mp \! X}$ subspace, the relative mode (red curve) is excited.
These trajectories are insensitive to $x_\mathrm{eq}\neq0$ as the $\ket{\pm X}$ are all eigenstates of the carrier interaction, which can manifest itelf as a state-dependent phase.
However, it turns out that $\Phi$ is also insensitive to the value of $x_\mathrm{eq}$ since the accumulated phase due to the carrier interaction from the first two pulses is removed by the second two pulses. The ultrafast gate is therefore robust to static offset fields.  The building blocks that make this gate can also be extended to include operations with co-trapped atomic ions whose motion is driven by lasers, allowing for hybrid applications requiring atom-molecule entanglement.

In summary, by using engineered electric-field gradients to drive transitions between electric-dipole-connected polar molecule internal and external states it is possible to construct a set of quantum logic gates that are largely independent of the motional state of the molecule.
Since the molecular qubits are controlled by microwave frequency voltages, this technique combines many of the desirable features of solid-state qubits with the long coherence times of trapped ion qubits.
The calculations presented here have used modest experimental parameters, e.g.\ $V_\mathrm{m} = 10$~V, $r_o = 0.5$~mm, $\omega_1 = 2\pi\cdot1$~MHz, and $x_\mathrm{eq} = 0.1 -10~\mu$m, which are routinely surpassed in many laboratories. 
Significant improvements of these parameters and the concomitant improvements in gate times and fidelities can be expected with additional techniques like superconducting microwave stripline resonators~\cite{Schuster2011}.
As a result, and in combination with the techniques of Ref.~\cite{Hudson2018,Campbell2020} and the possibility for robust encoding of qubits in rigid rotors~\cite{Albert2020}, polar molecular ions appear to be a promising system for constructing a large, scalable platform for quantum information science. 

\begin{acknowledgments}
This work was supported primarily by the U.S. Department of Energy, Office of Science, Basic Energy Sciences (BES), under Award No. DE-SC0019245, and partially by the ARO under Grant No. W911NF-19- 10297, the AFOSR under Grant No. FA9550-20-1-0323, and the NSF under Grants No. PHY-1806288, No. PHY- 1912555, and No. OMA-2016245.
\end{acknowledgments}

\bibliography{DipolePhonon}

\end{document}